\documentclass[11pt,a4paper]{article}
\usepackage{jcappub}
\usepackage{graphicx}
\usepackage{textcomp}
\usepackage{latexsym}
\usepackage{amsmath}
\usepackage{amssymb}

\def\lsim{\lower.5ex\hbox{$\; \buildrel < \over \sim \;$}}
\def\gsim{\lower.5ex\hbox{$\; \buildrel > \over \sim \;$}}

\def\pmb#1{\setbox0=\hbox{$#1$}%
\kern-.025em\copy0\kern-\wd0
\kern.05em\copy0\kern-\wd0
\kern-.025em\raise.0433em\box0}
\def\lsim{\lower.5ex\hbox{$\; \buildrel < \over \sim \;$}}
\def\gsim{\lower.5ex\hbox{$\; \buildrel > \over \sim \;$}}

\title{The role of axisymmetric flow configuration in 
the estimation of the analogue surface gravity and related 
Hawking like temperature}
\author[a]{Neven {Bili\'c,}}
\author[b,1]{Arpita {Choudhary,}}
\author[c,2]{Tapas K. {Das},} 
\author[d]{and Sankhasubhra {Nag}}

\affiliation[a]{Rudjer Bo\v{s}kovi\'{c} Institute, 10002 Zagreb, Croatia}
\emailAdd{bilic@thphys.irb.hr}

\affiliation[b]{University of Lucknow \note{Present Affiliation:Th\"{u}ringer Landessternwarte Tautenburg, Sternwarte 5, D-07778, Tautenburg, Germany}}\emailAdd{arpita@tls-tautenburg.de}

\affiliation[c]{Harish Chandra Research Institute Chhatnag Rd Jhunsi Allahabad 211 019 India \note{Presently on a long term sabbatical visit at 
S N Bose National Centre for Basic Sciences Salt Lake India}}\emailAdd{tapas@mri.ernet.in}

\affiliation[d]{Sarojini Naidu College for Women, Kolkata 700028, India}
\emailAdd{sankhasubhra\_nag@yahoo.co.in}

\abstract{For axially symmetric flow of dissipationless inhomogeneous fluid onto a 
non rotating astrophysical black hole under the influence of a generalized pseudo-Schwarzschild gravitational 
potential, we investigate the influence of the background flow 
configuration on determining the salient features of the corresponding 
acoustic geometry. The acoustic horizon for the aforementioned flow
structure has been located and the corresponding acoustic surface 
gravity $\kappa$ as well as the associated 
analogue Hawking temperature $T_{\rm AH}$ 
has been calculated {\it analytically}. The 
dependence of $\kappa$ on the flow geometry 
as well as on the nature of the back ground black hole space time 
(manifested through the nature of the pseudo-Schwarzschild 
potential used) has been discussed. Dependence of the value of 
$\kappa$ on various initial boundary conditions governing the 
dynamic and the thermodynamic properties of the background 
fluid flow has also been studied.} 

\begin{document}
\maketitle

\section{Introduction}
\label{sec1}
\noindent
Contemporary works in the field of analogue gravity phenomena have attracted significant 
attention in the community \cite{nvv02, cardoso05,blv05,
su07}.
Proper equivalence has been established between the physics of the propagating 
acoustic (and 
acoustic type) perturbations embedded in an inhomogeneous dynamical fluid system, and 
some kinematic features of the general theory of relativity. Such formalism has opened up 
the possibility of simulating various important features of the black hole space time
within the laboratory set up. 

Conventional works in this field, however, concentrates on 
the physical systems for which gravity like effects are realized as emergent phenomena. Such 
systems do not usually contain any source that produces active gravitational field in any 
form. In recent years, though, attempts have been made to study the analogue 
effects in strong gravity environment
\cite{das04, kst04, dgbd05, abd06, ncbr07, dbdg07, mm08, mach09,
pmdc12}.
The uniqueness 
of such systems lies in the fact that those are the only analogue models studied so far 
that simultaneously contain both kind of horizons, the gravitational as well as the acoustic -
allowing one to go for a close comparison between the actual and the analogue Hawking effect.

Till date, analogue effects in the axisymmetrically accreting black hole 
systems has been studied for the flow structure assumed to be in hydrostatic
equilibrium in the vertical direction. Two other configurations for the 
axially symmetric black hole accretion are also possible. Acccretion flow in conical 
equilibrium \cite{az81} and `flat disk' kind of flow with constant flow 
thickness have also been studied in the literature
(see, e.g., \cite{kfm98} for detail review). 
For pseudo-Schwarzschild gravitational potential
\footnote{In order to optimize between the easy handling of the Newtonian framework 
of gravity and more rigorous and non-tractable complete general relativistic 
description of the strong gravity space time, four different `modified' Newtonian 
`black hole potentials' have been introduced in the literature which are 
commonly known as pseudo-Schwarzschild potentials, see, e.g., 
\cite{abn96} , and \cite{das02}  for further detail about such 
potentials.}, the aforementioned three different flow structures have quite 
recently been studied in great detail (\cite{nard12}). to reveal 
various critical phenomena and stability properties of such flows. 

In our present work, we study the analogue effects in accreting 
black hole systems under the influence of pseudo-Schwarzschild 
potentials for three
different flow configurations - flow in conical and in vertical equilibrium and 
for constant thickness (height) flow. The main motivation behind this work 
is to study the dependence of the salient features of the acoustic geometry on the 
background geometry 
realized through the aforementioned three different flow configurations. To 
accomplish such task, we will calculate the acoustic surface gravity for same set of
accretion parameters but for all possible flow configurations in a 
pseudo-Schwarzschild 
potentials. 

\section{Acoustic surface gravity for classical analogue systems}
\noindent
Classical analogue gravity systems (alternatively, the classical `black hole 
analogues`) are fluid dynamical analogue of black holes in general 
relativity. Such analogue may occur when a 
small linear perturbation propagates through a dissipation less 
inhomogeneous barotropic transonic fluid at finite temperature. 
The corresponding acoustic metric, which specifies the geometry in which the perturbation propagates, may be constructed in 
terms of the flow variables defining the unperturbed background 
continuum. The transonic surface acts as acoustic horizon - a null 
hypersurface with acoustic null geodesics, the phonons, as its 
generators. The acoustic black hole horizon, which resembles the 
black hole event horizons in many ways, may form at the regular 
transonic point of the fluid, whereas an acoustic white hole horizon may 
be formed at the hypersurface where the fluid makes a 
discontinuous sonic transition, e.g., through a stationary shock
\cite{blsv04, abd06}. 

In his pioneering work Unruh \cite{unr81} introduced the concept of acoustic geometry 
inside a supersonic fluid and  demonstrated that an analogue surface 
gravity $\kappa$
may be associated with an acoustic black hole type event horizon, and one 
of the most interesting aspects of the acoustic horizon is to emit the 
Hawking type radiation of thermal phonons. Such acoustic Hawking 
radiation may be characterized by a analogue Hawking\footnote{Hereafter 
the phrases `acoustic' and `analogue' will be used synonymously for 
the sake of brevity.} temperature 
$T_{\rm AH}=\frac{{\hbar}{\kappa}}{2{\pi}}$. In Unruh's original approach,
the acoustic surface gravity could be associated with the component 
of the bulk velocity of the 
flow normal to the acoustic horizon
$u_{\perp}$ and the speed of propagation of the acoustic perturbation 
$c_s$ as
\begin{equation}
\kappa~{\propto}\left(\frac{1}{c_s}
\frac{{\partial}{u_{\perp}^2}}{{\partial}{\eta}}
\right)_{r_h}
\label{anal1}
\end{equation}
where $\frac{\partial}{\partial{\eta}}=\eta^{\mu}{\partial}_{\mu}$ 
represents the space derivative taken along the normal to the 
acoustic horizon, and every quantities in the 
eq. (\ref{anal1}) has been evaluated at the location of the 
acoustic horizon $r_h$. 

In eq. (\ref{anal1}), however, the sound speed was assumed to be a position 
independent constant. Unruh's work was followed by 
several other important contributions
(\cite{jacobson91, unr95, visser98, jacobson99} , to name a 
few)\footnote{It is interesting to note that the concept of the 
acoustic geometry inside a transonic fluid was first realized by 
\cite{mon80}  while studying the stability properties of 
the relativistic spherical accretion onto astrophysical black holes.}
Visser \cite{visser98}  implemented the additional contribution due to the position 
dependent sound speed and obtained  a modified expression for the surface gravity
\begin{equation}
\kappa~{\propto}
\left[c_s\frac{\partial}{\partial{\eta}}
\left(c_s-u_\perp\right)\right]_{ r_h}
\label{anal2}
\end{equation}
The acoustic horizon  is a surface  defined by  the 
equation \cite{visser98} 
\begin{equation}
u_{\perp}^2-c_s^2=0
\label{anal3}
\end{equation}
for the stationary background flow configuration. Equation  (\ref{anal3}) basically
states that the acoustic horizons is a transonic 
surface. The supersonic region of a transonic flow defines
the acoustic ergo region.

The concept of  acoustic geometry has been extended  
to a relativistic fluid flow  in a general background space time \citep{bilic99}.
For an ideal barotropic fluid, the relativistic 
Euler equation and the equation of continuity obtained from 
the energy momentum conservation  can be 
linearized in order  to obtain the wave equation for the propagating 
perturbation in analogue curved space time with the corresponding acoustic metric. The generalized 
form of the acoustic surface gravity turns out to be 
\begin{equation}
\kappa=\left|\frac{\sqrt{{\chi^\mu}{\chi_\mu}}}{\left(1-{c_s}^2\right)}
\frac{\partial}{\partial{\eta}}\left(u_\perp-{c_s}\right)\right|_
{ r_h}
\label{anal4}
\end{equation}
where $\chi^\mu$ is the Killing field which is null on the corresponding
acoustic horizon. 
The algebraic expression corresponding to the $\sqrt{\chi^\mu{\chi_\mu}}$
may thus be evaluated once the background stationary metric governing the 
fluid flow as well as the propagation of the perturbation in a specified 
geometry with well posed boundary conditions are 
realized. It is worth mentioning that the generalized form 
for $\kappa$ as defined in eq. (\ref{anal4}) can further be reduced to 
its Newtonian/semi Newtonian counterpart depending on the nature of the 
gravitational potential describing the background fluid motion. 

\section{Acoustic surface gravity for axisymmetric black hole 
accretion}
\noindent
 From (\ref{anal2}) and (\ref{anal4}) it is clear that, in order 
to find the acoustic surface gravity $\kappa$ 
for the Newtonian as well as for the general relativistic acoustic 
geometry, it is sufficient to calculate
 the location $r_h$ of the acoustic horizon,
the sound speed $c_s$ of the small 
linear perturbation and its normal space gradient $dc_s/d{\eta}$, 
as well as the  the normal (to the acoustic horizon) 
component of the flow velocity $u_\perp$ and its normal space 
gradient  $du_\perp/d{\eta}$, evaluated on the acoustic horizon. 

For transonic accretion onto astrophysical black hole, one thus needs to 
consider the Euler equation and the equation of continuity for a specific 
symmetry of the problem. The Euler and the continuity equation may then
be linearized (for a 
general linearization scheme and its application to the study of 
axisymmetric black hole accretion for three different flow geometries see  \citep{nard12}) in order
to construct the corresponding acoustic metric to 
specify the relevant acoustic geometry. The structure of the 
stationary background fluid flow may be provided by the stationary 
solution of the Euler and the continuity equations. For a 
certain set of values of the initial conditions describing 
the transonic accretion flow governed by certain barotropic equation of 
state, it may be possible to find the location of the saddle type 
transonic point, which is identical to the acoustic black hole 
horizon (see~\cite{dgbd05,
bilic99, abd06,
dbdg07} ). The quantities
$\left[u_\perp,c_s,du_\perp/dr,dc_s/dr\right]_{ r_{\rm transonic}}$
subject to a gravitational 
field which governs the accretion, may then be evaluated to estimate the acoustic surface 
gravity using eq. (\ref{anal4}). The quantity
$\sqrt{\chi^\mu{\chi_\mu}}$
evaluated at the acoustic horizon is a  function of 
${ r_h}={ r_{\rm transonic}}$ (as will be 
demonstrated in subsequent paragraphs)   depending on initial conditions.

In general, we consider the equatorial slice of the
axisymmetric gravitating accretion of 
hydrodynamic fluid onto non rotating black holes. The gravitational field
of the black hole is assumed to be described by certain 
pseudo-Schwarzschild potentials. The Schwarzschild radius 
$r_g=2GM_{\rm BH}/c^2$ is used to scale the radial distance,
whereas all velocities involved are scaled by the velocity of light in 
vacuum $c$, $G=c=1$ has been used. Accretion is assumed to 
possess finite radial velocity $u$ commonly known as the
`advective velocity' in the accretion literature. 
Considering $v$ to be the magnitude of the three velocity, $u$
is the component of three velocity perpendicular to the set of timelike
hypersurfaces $\left\{\Sigma_v\right\}$ defined by $v^2={\rm constant}$.
The advective velocity $u$ is thus perpendicular to the 
acoustic horizon and hence $u$ is identical with $u_\perp$. Hereafter we 
drop the subscript $\perp$ in $u_\perp$ and simply use $u$ instead. 

The low angular momentum sub-Keplerian advective
inviscid flow will 
be considered where the specific flow angular momentum $\lambda$ will 
be assumed to be a position independent constant. Viscous transport
of angular momentum will not be taken into account since 
close to the
black hole, the infall time scale for the highly supersonic flow is
rather small compared to the corresponding viscous time scale
(see, e.g., \cite{das02, dpm03, dcz12}  and references therein for further detail). 
Also for advective
accretion, large radial velocity at a larger distances are the consequence of
the small rotational energy of the flow
\cite{bi91,
ib97, pb03} . The 
corresponding angular velocity $\Omega$ may be defined as 
(\cite{dcz12}  and references therein) 
\begin{equation}
\Omega==-\frac{g_{t\phi}+\lambda{g}_{tt}}{{g_{\phi{\phi}}+\lambda{g}_{t{\phi}}}}
\label{anal5}
\end{equation}
where $g_{ij}$  are the metric components.

The corresponding surface gravity $\kappa$ can now be expressed as
\begin{equation}
\kappa=
\left|
\sqrt{\frac{\chi^\mu{\chi_\mu}}{-g_{rr}}}
\frac{1}{1-c_s^2}
\left[\frac{d}{d{r}}\left(u-{c_s}\right)\right]\right|_
{r_h}
\label{anal6}
\end{equation}
where $\chi^\mu=\xi^\mu+\Omega{\varsigma^\mu}$ and the Killing vectors
$\xi^\mu$ and $\varsigma^\mu$ are the generators of the temporal 
and the axial symmetry group. The norm of the Killing vector $\chi_\mu$
may be computed as 
\begin{equation}
\sqrt{\chi^\mu{\chi_\mu}}
=\left(g_{tt}+2\Omega{g_{t\phi}}+\Omega^2{g_{\phi\phi}}\right)^{\frac{1}{2}}=
\frac{{\Sigma}{\Lambda}}{g_{\phi{\phi}}+\lambda{g_{t\phi}}}
\label{anal7}
\end{equation}
where
\begin{eqnarray}
\Sigma^2=g^2_{t{\phi}}-g_{tt}g_{\phi{\phi}}\\
\nonumber
\Lambda^2=\left(g_{tt}+2\lambda{g_{t\phi}}+\lambda^2{g_{\phi\phi}}\right)
\label{anal8}
\end{eqnarray}
In the Newtonian limit 
\begin{equation}
g_{tt}=1+2\phi,~
g_{{\phi}{\phi}}=-r^2,~
g_{rr}=-1,~
g_{t\phi}=0
\label{anal9}
\end{equation}
where $\Phi$ is the pseudo potential.
Hence the acoustic gravity for axisymmetric black hole accretion under the 
influence of the pseudo-Schwarzscild potential becomes
\begin{equation}
\kappa=
\left|
\sqrt{
\left(1+\Phi\right)
\left(1-\frac{\lambda^2}{r^2}-2\Phi\frac{\lambda^2}{r^2}\right)
}
\left(
\frac{1}{1-c_s^2}
\left[
\frac{du}{dr}-\frac{dc_s}{dr}
\right]
\right)
\right|_{r_h}
\label{anal10}
\end{equation}

Our main task now boils down to the evaluation of $c_s$,
$dc_s/dr$ and $du/dr$ on the acoustic horizon for the
pseudo-Schwarzschild axisymmetric black hole accretion
in three different flow configurations as mentioned in 
previous sections.

\section{Generalized transonic accretion in three different flow geometries}
The governing equations describing the dynamics of axially symmetric 
pseudo-Schwarzschild inviscid hydrodynamics accretion are the equation for the 
conservation of linear momentum (the Euler equation):
\begin{equation}
\frac{{\partial}}{{\partial}t}u(r,t)+
u(r,t)\frac{{\partial}}{{\partial}r}u(r,t)+
\frac{1}{\rho(r,t)}\frac{{\partial}p(r,t)}{{\partial}r}-
\frac{\lambda^2}{r^3}+
\Phi'=0
\label{eq1}
\end{equation}
where $u,\rho$ and $p$, being the dynamical flow velocity, the fluid density and the 
pressure, respectively, are functions of both $r$ and $t$. 
$\Phi$ may be taken as any one of the 
following four pseudo-Schwarzschild potentials:
\begin{subequations}\label{phi}
\begin{eqnarray}
\Phi_{1} &=& -\frac{1}{2(r-1)},\\
\Phi_{2} &=& -\frac{1}{2r}\left[1-\frac{3}{2r}+12{\left(\frac{1}{2r}\right)}^2\right],\\
\Phi_{3} &=& -1+{\left(1-\frac{1}{r}\right)}^{\frac{1}{2}},\\
\Phi_{4} &=& \frac{1}{2}{\rm ln}{\left(1-\frac{1}{r}\right)}
\end{eqnarray}  
\end{subequations}

The potential $\Phi_1$ and $\Phi_2$ have been introduced in
\cite{pw80}  and  \cite{nw91} , respectively,
whereas $\Phi_3$ and $\Phi_4$ have been introduced in 
\cite{abn96} .

The mass conservation equation 
(the continuity equation) can be written as:
\begin{equation}
\frac{{\partial}}{{\partial}t}\rho(r,t)+
\frac{\partial}{{\partial}{r}}\left[\rho(r,t){u(r,t)rH}\right]=0
\label{eq3}
\end{equation}
The quantity $H$ is the flow thickness which is different in three different 
flow configurations. For the simplest possible flow configuration, the flow 
thickness (i.e., the disc height) is constant, and hence $H$ is not a function 
of the radial distance. In its next variant, the flow can have 
a conical structure (see~\cite{az81} ) where $H$ is directly 
proportional to the radial distance as $H=Ar$, where the geometric 
constant $A$ depends on the solid angle subtended by the flow.
For hydrostatic equilibrium in vertical direction
\cite{mkfo84, fkr02}, the flow thickness
can have a rather complex dependence on the radial distance and on  the local adiabatic sound speed   \citep{das02,dpm03}
 defined as 
$c_s=\sqrt{\left(\frac{\partial{p}}{\partial{\rho}}\right)}$.
The barotropic equation of state will in general be used to 
describe the accretion flow in this work. A polytropic equation 
of state $p=K{\rho}^\gamma$ will be used, whereas the isothermal 
flow will be governed by the equation $p=\frac{\rho{\kappa_B}T}{{\mu}m_H}$.
The quantities $K,\gamma,\kappa_B,T,\mu$ and $m_H$ are the entropy per 
particle, the Boltzmann constant, the isothermal flow temperature,
the reduced mass and the mass of the Hydrogen atom, respectively. 
With the help of the equation of states, and the specified 
radial dependence of the flow thickness, we can find 
stationary solutions of the Euler and the continuity equations and draw the Mach number versus the radial distance phase 
portrait for the integral flow solutions to obtain the detail 
information about the location of the acoustic horizon as well 
as the horizon related quantities.

\subsection{Polytropic Accretion}
\label{polytropic_accretion}
For polytropic equation of state, the integral solution of the 
stationary part of the Euler equation provides the energy first 
integral of motion of the following form:
\begin{equation}
{\cal E}=\frac{u^2}{2}+\frac{c_s^2}{\gamma-1}+\frac{\lambda^2}{2r^2}+\Phi
\label{eq5}
\end{equation}
The conserved specific energy ${\cal E}$ does not depend on the 
flow configuration for obvious reason. Since $r$ dependence of $H$ 
varies for different flow geometries, the integral solution of the 
continuity equation, which is another first integral of motion and 
is referred to as the mass accretion rate, will be different for three
different accretion configurations. Expressions for the 
mass accretion rate can be obtained as:
\begin{subequations}\label{eq6}
 \begin{eqnarray}
{\dot M}_{\rm CH}=\rho{u}rH_c\\
{\dot M}_{\rm CM}=\Theta{\rho}ur^2\\
{\dot M}_{\rm VE}=
\sqrt{\frac{1}{\gamma}}uc_s{\rho}r^{\frac{3}{2}}
\left(\Phi'\right)^{-\frac{1}{2}}
\end{eqnarray} 
\end{subequations}
where the subscript CH,CM and VE stands for the 
flow with constant height (CH), in conical model
(CM), and in vertical equilibrium (VE), and implies that 
the respective algebraic equations are to be solved
to obtain the critical point for the corresponding flow geometries.
The quantity $H_c$ is the constant disc height and $\Theta$ is the solid angle 
sustained by the flow. The mass accretion rate for flow in vertical equilibrium not 
only depends on the matter geometry (through the radial dependence of $H$), but 
also the information about the space time geometry is encrypted in 
${\dot M}_{\rm VE}$ through the explicit appearance of the 
derivative of the pseudo-Schwarzschild potential. One defines the entropy accretion 
rate as (\cite{az81,blaes87}):
\begin{equation}
{\dot {\cal M}}=
{\dot M}\gamma^{\frac{1}{\gamma-1}}K^{\frac{1}{\gamma-1}}
\label{eq7}
\end{equation}
Substitution of ${\dot M}$ from eq. (\ref{eq6}) in the above equation provides the 
expression for ${\dot {\cal M}}$ in terms of 
the adiabatic sound speed, radial distance, and the dynamical flow velocity. 
The space gradient of the sound speed as well as the dynamical velocity 
for various flow geometries can be obtained by differentiating 
eq. (\ref{eq5}) and eq. (\ref{eq7}):
\begin{subequations}\label{eq8}
\begin{eqnarray}
\left(\frac{dc_s}{dr}\right)_{\rm CH}&=&(1-\gamma)\frac{c_s}{u}\left(\frac{1}{2}\frac{du}{dr}+\frac{u}{2r}\right) \\
\left(\frac{dc_s}{dr}\right)_{\rm CM}&=&(1-\gamma)\frac{c_s}{u}\left(\frac{1}{2}\frac{du}{dr}+\frac{u}{r}\right) \\
\left(\frac{dc_s}{dr}\right)_{\rm VE}&=&\left(\frac{1-\gamma}{1+\gamma}\right)\frac{c_s}{u}\left[\frac{du}{dr}+\frac{u}{2}\left(\frac{3}{r}-\frac{\Phi''(r)}{\Phi'(r)}\right)\right] 
\end{eqnarray}  
\end{subequations}
\begin{subequations}\label{eeq9}
\begin{eqnarray}
\left(\frac{du}{dr}\right)_{\rm CH}&=&\frac{u\left(\frac{c_s^2}{r}+\frac{\lambda^2}{r^3}-\Phi'(r)\right)}{\left(u^2-c_s^2\right)} \\
\left(\frac{du}{dr}\right)_{\rm CM}&=&\frac{u\left(\frac{2c_s^2}{r}+\frac{\lambda^2}{r^3}-\Phi'(r)\right)}{\left(u^2-c_s^2\right)} \\
\left(\frac{du}{dr}\right)_{\rm VE}&=&\frac{u\left[\frac{c_s^2}{(1+\gamma)}\left(\frac{3}{r}-\frac{\Phi''(r)}{\Phi'(r)}\right)+\frac{\lambda^2}{r^3}-\Phi'(r)\right]}{\left(u^2-\frac{2}{1+\gamma}c_s^2\right)}  
\end{eqnarray}  
\end{subequations}

The 
critical point conditions may be
obtained by simultaneously making the numerator and the denominator of
eq. (\ref{eeq9}) vanish, and the aforementioned critical point
conditions may thus be expressed as:
\begin{subequations}\label{eq10}
\begin{eqnarray}
\left(u\right)_{r_c}=\left({c_s}\right)_{rc} \\
\left({c_s}\right)_{r_c}=\sqrt{r_c\Phi'(r_c)-\frac{\lambda^2}{r_c^2}}
\end{eqnarray}  
\end{subequations}

\begin{subequations}\label{eq11}
\begin{eqnarray}
\left(u\right)_{r_c}=\left({c_s}\right)_{rc} \\
\left({c_s}\right)_{r_c}=\sqrt{\frac{r_c\Phi'(r_c)}{2}-\frac{\lambda^2}{2r_c^2}}
\end{eqnarray}
\end{subequations}

\begin{subequations}\label{eq12}
\begin{eqnarray}
\left(u\right)_{r_c}=\sqrt{\frac{2}{\gamma+1}}\left({c_s}\right)_{rc} \\
\left({c_s}\right)_{rc}=\sqrt{(\gamma+1)\frac{\left[\Phi'(r_c)-\frac{\lambda^2}{r_c^3}\right]}{\left[\frac{3}{r_c}-\frac{\Phi''(r_c)}{\Phi'(r_c)}\right]}}
\end{eqnarray}  
\end{subequations}

The critical point conditions for the constant height flow, the conical model flow and 
flow in vertical equilibrium are stated in eq. (\ref{eq10}), (\ref{eq11}) and (\ref{eq12}) 
respectively. 
Linearizing the Euler and the continuity equation for accretion in hydrostatic equilibrium 
in the vertical direction, one can show that 
the linear perturbation propagates with the speed 
$\sqrt{\frac{2}{1+\gamma}}c_s$ instead of $c_s$. We thus define 
$\sqrt{\frac{2}{1+\gamma}}c_s$ to be the `effective' sound speed for such flow.
This happens because for flow in vertical equilibrium the expression for the 
flow thickness contains the adiabatic sound speed $c_s=\gamma{p}/{\rho}$.

For a set of fixed values of 
$\left[{\cal E},\lambda,\gamma\right]$, the location of the critical 
point for a particular flow model can be obtained by 
substituting the corresponding critical point condition (as expressed in eq. (\ref{eq10} - \ref{eq12})
in the energy first integral eq.~(\ref{eq5}) for a particular pseudo-Schwarzschild black hole potential. Once these expressions are 
substituted, the energy first integral becomes an algebraic expression of $r_c$. Exact 
value of $r_c$ for the constant height flow, 
flow in conical model, and in hydrostatic equilibrium 
can thus be obtained by solving the following equations
\begin{subequations}\label{eq13}
\begin{eqnarray}
{\cal E}_{\rm CH}-\frac{1}{2}
\left(\frac{\gamma +1}{\gamma -1}\right)
\left[r_{{c}} \left(\Phi'\right)_{r_c}
- \frac{\lambda^2}{r_{{c}}^2} \right] 
- \Phi (r_{{c}}) 
- \frac{\lambda^2}{2 r_{{c}}^2}=0 \\
{\cal E}_{\rm CM}-
\frac{1}{4} 
\left(\frac{\gamma +1}{\gamma -1}\right)
\left[r_{{c}} \left(\Phi'\right)_{r_c}
- \frac{\lambda^2}{r_{{c}}^2} \right] 
- \Phi (r_{{c}}) 
- \frac{\lambda^2}{2 r_{{c}}^2} = 0 \\
{\cal E}_{\rm VE}-
\frac{2 \gamma}{\gamma -1} 
\left[r_{{c}} \Phi'(r_{{c}})
- \frac{\lambda^2}{r_{{c}}^2} \right] \left[3 - r_{{c}}
\left(\frac{\frac{d^2\Phi}{dr^2}}{\Phi'}\right)_{r_c}
\right]^{-1} - \Phi (r_{{c}}) 
- \frac{\lambda^2}{2 r_{{c}}^2} =0
\end{eqnarray}  
\end{subequations}

The exact location of $r_c$ can be evaluated once the astrophysically relevant range
of $\left[{\cal E},\lambda,\gamma\right]$ can be realized. One can argue 
\cite{pmdc12}  that the relevant values in the parameter space 
$\{ {\cal E},\lambda,\gamma \} $ can be set as
$\left[1{\lsim}{\cal E}{\lsim}2,0<\lambda{\le}2,4/3{\le}\gamma{\le}5/3\right]$.

A solution of eq. (\ref{eq13}) may  exhibit 
either
one (saddle type), or three (one centre type flanked by two saddle
type) critical points depending on the chosen set of parameters $\left[{\cal E},\lambda,\gamma\right]$
used.
Certain $\left[{\cal E},\lambda,\gamma\right]$ $_{\rm mc}{\subset}$
$\left[{\cal E},\lambda,\gamma\right]$ thus provides the
multi critically in  accretion 
solutions, where the subscript `mc' stands for `multi critical'.
The acoustic horizon are thus the collection of the
`sonic' points where the radial Mach number becomes unity.
Such a horizon is located on the combined integral solution of
eq. (\ref{eq8}) and eq. (\ref{eeq9}). For inviscid flow, a physically
acceptable transonic solution which  passes through a saddle type
sonic point can be realized. Such a solution would be an example which confirms the hypothesis that every saddle type critical
point is accompanied by its sonic point but no centre type critical point
has its sonic counterpart. For an axisymmetric configuration, in all three geometries discussed in this work,
a multi-critical flow is thus a theoretical
abstraction where three critical points (out of which one is 
always a centre type, through which the integral solution can never pass)
are obtained as a mathematical
solution of the energy conservation equation (through the critical point
condition), whereas a multi-transonic flow is a realistic
configuration where accretion solution passes through two different saddle
type sonic points.
One should, however, note
that a smooth accretion solution can never encounter more than one regular
sonic points, hence no continuous transonic solution exists which passes
through two different acoustic horizons. The only way the multi transonicity
could be realized as a combination of two different otherwise smooth solutions
passing through two different saddle type critical (and hence sonic) points and
are connected to each other through a discontinuous shock transition.
Such a shock has to be stationary and will be located in between two sonic
points.
For a specific
$\left[{\cal E},\lambda,\gamma\right]_{\rm No~Shock}{\subset}$
$\left[{\cal E},\lambda,\gamma\right]_{\rm mc}$,
three
critical points (two saddle embracing a centre one)
are routinely obtained but no stationary shock
forms. Hence no multi transonicity is observed even if the flow is
multi-critical, and real physical accretion solution can have access only
to the outer type saddle point out of the two. Thus multi
critical accretion and multi transonic accretion are not topologically
isomorphic in general. A true multi-transonic flow can only be
realized for
$\left[{\cal E},\lambda,\gamma\right]_{\rm Shock}{\subset}$
$\left[{\cal E},\lambda,\gamma\right]_{\rm mc}$,
if the criteria for forming a
standing shock are met (for details about such shock formation and 
related multi-transonic shocked flow topologies see \cite{cd01,das02,dpm03}).
For mono-transonic flow, one can have only one acoustic horizon on which the
related surface gravity may be evaluated. For multi-transonic 
shocked accretion, however, one can have two acoustic horizons 
(at the inner and the outer saddle type critical point) and can 
calculate the corresponding two different values of the 
acoustic surface gravity. We show this in subsequent sections.

Once the critical point is located, 
the critical derivatives of the sound speed $\left(\frac{dc_s}{dr}\right)_{r_c}$ and of the flow velocity
$\left(\frac{du}{dr}\right)_{r_c}$, evaluated at the critical
point ${r_c}$
(which coincides with the location of the acoustic horizon 
${r_h}$), can be obtained for various flow models
by applying L' Hospital's rule to the numerator and the
denominator of eq. (\ref{eeq9})
\begin{subequations}
 \begin{eqnarray}
\left|\left(\frac{du}{dr}\right)_{\rm CH}\right|_{r_c}&=&\frac{1}{r_c}\left(\frac{1-\gamma}{1+\gamma}\right)\sqrt{{r_c\Phi'(r_c)}-\frac{\lambda^2}{r_c^2}}\\ \nonumber &&\pm\sqrt{\frac{1}{r_c^2}\left(\frac{1-\gamma}{1+\gamma}\right)^2\left({r_c\Phi'(r_c)}-\frac{\lambda^2}{r_c^2}\right)-\frac{\left(\frac{\gamma}{r_c^2}(r_c\Phi'(r_c)-\frac{\lambda^2}{r_c^2})+\frac{3\lambda^2}{r_c^4}+\Phi''(r_c)\right)}{\sqrt{{r_c\Phi'(r_c)}-\frac{\lambda^2}{r_c^2}}}} 
\label{dudrc-adiabatic-constant-height}
\end{eqnarray}
\begin{eqnarray}
\left|\left(\frac{du}{dr}\right)_{\rm CM}\right|_{r_c}&=&\frac{2}{r_c}\left(\frac{1-\gamma}{1+\gamma}\right)\sqrt{\frac{r_c\Phi'(r_c)}{2}-\frac{\lambda^2}{2r_c^2}}\\ \nonumber &&\pm\sqrt{\frac{4}{r_c^2}\left(\frac{1-\gamma}{1+\gamma}\right)^2\left(\frac{r_c\Phi'(r_c)}{2}-\frac{\lambda^2}{2r_c^2}\right)-\frac{\left(\frac{2(2\gamma-1)}{r_c^2}(\frac{r_c\Phi'(r_c)}{2}-\frac{\lambda^2}{2r_c^2})+\frac{3\lambda^2}{r_c^4}+\Phi''(r_c)\right)}{(1+\gamma)\sqrt{\frac{r_c\Phi'(r_c)}{2}-\frac{\lambda^2}{2r_c^2}}}} 
\label{dudrc-adiabatic-conical-model}
\end{eqnarray}
\begin{eqnarray}
\left|\left(\frac{du}{dr}\right)_{\rm VE}\right|_{r_c}&=&2u_c\left(\frac{\gamma-1}{8\gamma}\right)\left[\frac{3}{r_c}+\frac{\Phi'''(r_c)}{\Phi'(r_c)}\right]\\ \nonumber &\pm& \sqrt{\frac{\gamma+1}{4\gamma}}\left[u_c^2\frac{\gamma-1}{\gamma+1}\frac{\gamma-1}{4\gamma}\left(\frac{3}{r_c}+\frac{\Phi''(r_c)}{\Phi'(r_c)}\right)^2\right.  \\ \nonumber  &-& \left. u_c^2\frac{1+\gamma}{2}\left(\frac{\Phi'''(r_c)}{\Phi'(r_c)}-\frac{2\gamma}{(1+\gamma)^2}\left(\frac{\Phi'''(r_c)}{\Phi'(r_c)}\right)^2\right.\right. \\ \nonumber &+&\left.\left.\frac{6(\gamma-1)}{\gamma(\gamma+1)^2}\frac{\Phi''(r_c)}{\Phi'(r_c)}-\frac{6(2\gamma-1)}{\gamma^2(\gamma+1)^2}\right) -  \Phi''(r_c)+\frac{3\lambda^2}{r_c^4}\right]^{1/2} 
\label{dudrc-adiabatic-vertical-equilibrium}
\end{eqnarray} 
\end{subequations}
The quantity $u_c$ in (\ref{eq11}) may 
be substitutted from eq. (\ref{eq12}). 


The acoustic surface gravity $\kappa$ as defined in eq. (\ref{anal10}) may now be 
evaluated for various space time geometries for adiabatic accretion. The 
location of the acoustic horizon (the critical point $r_c$) and 
$\left[u,c_s,dc_s/dr,du/dr\right]_{r_c}$ can be evaluated 
as a function of the initial boundary conditions as defined by the 
parameters $\left[{\cal E},\lambda,\gamma\right]$ 
for the adiabatic flow and $\left[T,\lambda\right]$ for the 
isothermal flow (the details of the 
calculation of $\kappa$ for the isothermal accretion will be presented in the next section) for a fixed flow 
geometry in all four pseudo potentials as well as under the 
influence of a particular pseudo potentials in all three different 
flow geometries as considered in this work. 

\subsection{Isothermal Accretion}
For isothermal flow, the integral solution of the time independent Euler equation 
provides the following first integral of motion
\begin{equation}
\frac{u^2}{2} + c_{\mathrm s}^2 \ln \rho 
+ \frac{\lambda^2}{2 r^2} + \Phi (r) = {\rm Constant}
\label{iso1}
\end{equation}
Obviously, this constant of motion can not be identified with the 
specific energy of the flow. The isothermal sound speed is proportional to $T^\frac{1}{2}$. The mass accretion rate,
another first integral of motion of the accreting system of aforementioned kind, may 
be obtained for three different flow geometries as
\begin{subequations}\label{iso2}
\begin{eqnarray}
{\dot M}_{\rm CH}^{\rm iso}=\rho{u}rH_c\\
{\dot M}_{\rm CM}^{\rm iso}=\Theta{\rho}ur^2\\
{\dot M}_{\rm VE}^{\rm iso}=
c_s\rho{u}r^{\frac{3}{2}}\left({\Phi'}\right)^{-\frac{1}{2}}
\end{eqnarray}  
\end{subequations}
The space gradient of the velocities for these three models comes out to be
\begin{subequations}\label{iso3}
\begin{eqnarray}
\left(\frac{du}{dr}\right)_{\rm CH}^{\rm iso}&=&\frac{u\left(\frac{c_s^2}{r}-\Phi'(r)+\frac{\lambda^2}{r^3}\right)}{\left(u^2-c_s^2\right)} \\
\left(\frac{du}{dr}\right)_{\rm CM}^{\rm iso}&=&\frac{u\left(\frac{2c_s^2}{r}-\Phi'(r)+\frac{\lambda^2}{r^3}\right)}{\left(u^2-c_s^2\right)} \\
\left(\frac{du}{dr}\right)_{\rm VE}^{\rm iso}&=&\frac{u\left[\frac{c_s^2}{2}\left(\frac{3}{r}-\frac{\Phi''(r)}{\Phi'(r)}\right)-\Phi'(r)+\frac{\lambda^2}{r^3}\right]}{\left(u^2-c_s^2\right)} 
\end{eqnarray}  
\end{subequations}
which provides the following critical point conditions
\begin{equation}
\left(u\right)_{r_c}=
\left(c_s\right)_{r_c}=
\sqrt{\frac{\kappa_B}{{\mu}m_H}}T^{\frac{1}{2}}=
\sqrt{r_{\mathrm{c}} \left[\Phi'\right]_{r_c}
- \frac{\lambda^2}{r_{\mathrm{c}}^2}}
\label{iso4}
\end{equation}
\begin{equation}
\left(u\right)_{r_c}=
\left(c_s\right)_{r_c}=
\sqrt{\frac{\kappa_B}{{\mu}m_H}}T^{\frac{1}{2}}=
\sqrt{\frac{1}{2}\left(r_{\mathrm{c}} \left[\Phi'\right]_{r_c}
- \frac{\lambda^2}{r_{\mathrm{c}}^2}\right)}
\label{iso5}
\end{equation}
and
\begin{equation}
\left(u\right)_{r_c}=
\left(c_s\right)_{r_c}=
\sqrt{\frac{\kappa_B}{{\mu}m_H}}T^{\frac{1}{2}}=
\sqrt{2}\left(r_{\mathrm{c}} \left[\Phi'\right]_{r_c}
- \frac{\lambda^2}{r_{\mathrm{c}}^2} \right)^{\frac{1}{2}} 
\left(3 - r_{\mathrm{c}}
\left[\frac{\Phi''}{\Phi'}\right]_{r_c}
\right)^{-\frac{1}{2}}
\label{iso6}
\end{equation}
for flow with constant thickness (eq. \ref{iso4}), in conical equilibrium 
(eq. \ref{iso5}) and for flow in hydrostatic equilibrium in the vertical direction 
(eq. \ref{iso6}), respectively. A two parameter input $\left[T,\lambda\right]$
($T$ being the isothermal flow temperature), can solve eq. (\ref{iso4}) - (\ref{iso6}) to 
obtain the location of the acoustic horizon for three different flow configurations as mentioned 
above. The critical space gradient of the dynamical velocity as evaluated on the 
acoustic horizon are given by
\begin{subequations}
\begin{equation}
\left|\left(\frac{du}{dr}\right)^{\rm iso}_{\rm CH}\right|_{r_c}=\pm\frac{1}{\sqrt{2}}\sqrt{-\Phi''(r_c)-\left(\frac{c_s^2}{r_c^2}+\frac{3\lambda^2}{r_c^4}\right)}
\label{iso7}
\end{equation}
\begin{equation}
\left|\left(\frac{du}{dr}\right)^{\rm iso}_{\rm CM}\right|_{r_c}=\pm\frac{1}{\sqrt{2}}\sqrt{-\Phi''(r_c)-\left(\frac{2c_s^2}{r_c^2}+\frac{3\lambda^2}{r_c^4}\right)}
\label{iso8}
\end{equation}
and
\begin{equation}
\left|\left(\frac{du}{dr}\right)^{\rm iso}_{\rm VE}\right|_{r_c}=\pm\frac{1}{\sqrt{2}}\sqrt{\frac{c_s^2}{2}\left[\left(\frac{\Phi''(r_c)}{\Phi'(r_c)}\right)^2-\left(\frac{\Phi'''(r_c)}{\Phi'(r_c)}\right)\right]-\left(\Phi''(r_c)+\frac{3c_s^2}{2r_c^2}+\frac{3\lambda^2}{r_c^4}\right)}
\label{iso9}
\end{equation}  
\end{subequations}
Hence the acoustic surface gravity for isothermal 
accretion can be evaluated for three different flow models as a 
function of only two parameters, namely, the flow angular momentum $\lambda$ and 
the isothermal flow temperature $T$.

\section{Analytical calculation of the acoustic surface gravity and the 
corresponding analogue Hawking temperature}
In this section we calculate the surface gravity $\kappa$ for  a fluid gravitating in the Paczy\'nski and Wiita (1980) pseudo-Schwarzschild potential 
$\Phi_1=-\frac{1}{2\left(r-1\right)}$ for flow with constant height, in conical shape and in 
hydrostatic equilibrium in the vertical direction, for both the adiabatic as well as the
isothermal accretion, and will study the variation of $\kappa$ for three 
different flow geometries used. In addition, we calculate $\kappa$ for the same 
set of initial boundary conditions 
 for both the adiabatic and the isothermal accretion
under the influence of all four pseudo Schwarzschild potentials as defined by (\ref{phi}), for a flow in any of the three geometries 
mentioned before. Studying the acoustic surface gravity $\kappa$ as a function of $\Phi$ 
provides information about the dependence of  $\kappa$
 on the background space time geometry.

\subsection{Adiabatic Accretion}
In subsequent sections, we will calculate $\kappa$ for three different models using 
Paczy\'nski and Wiita potential \cite{pw80} for adiabatic accretion. 

\subsubsection{Accretion flow with constant thickness}
We start with the simplest flow configuration - axisymmetric flow with constant 
thickness. For such flow, the space gradient of the speed of sound and the flow velocity can be computed as:
\begin{subequations}\label{eq14} 
\begin{eqnarray}
\frac{dc_{s}}{dr}=\frac{c_{s}(1-\gamma)}{2}\left[\frac{1}{r}+\frac{1}{u}\frac{du}{dr}\right]  \\
\frac{du}{dr}= \frac{u \left[  \frac{c_{s}^2}{r}+
\frac{\lambda^2}{r^3} - \frac{1}{2(r-1)^2}  \right] }{(u^2-c_{s}^2)}  
\end{eqnarray}  
\end{subequations}
The corresponding critical point conditions can thus be obtained as:
\begin{equation}
\left(c_{s}\right)_{r_c}= 
\left(u\right)_{r_c} = 
\sqrt{\left[ \frac{r_{c}}{2(r_{c}-1)^{2}} - \frac{\lambda^{2}}{r_{c}^2} \right]_{r_c}} 
\label{eq15}
\end{equation}
By substituting the above condition into the equation for 
the energy first integral (\ref{eq5}), 
a fourth degree polynomial of $r_c$ of the 
following form can be obtained:
\begin{eqnarray}\label{eq16}
r_{c}^4+\Gamma_{1}r_{c}^3+\Gamma_{2}r_{c}^2+\Gamma_{3}r_{c}+\Gamma_{4}=0 \\ \nonumber
{\rm where} \\ \nonumber
\Gamma_{1}=\frac{(\gamma-3)-8\mathcal{E}(\gamma-1)}{4\mathcal{E}(\gamma-1)} ; 
~ \Gamma_{2}=\frac{(2\mathcal{E}-1)(\gamma-1)+2\lambda^2}{2\mathcal{E}(\gamma-1)}  \\ \nonumber
\Gamma_{3}=\frac{-2\lambda^2}{\mathcal{E}(\gamma-1)} ; 
~ \Gamma_{4}=\frac{\lambda^2}{\mathcal{E}(\gamma-1)}
\end{eqnarray}
The location of the acoustic horizon in terms of 
$\left[{\cal E},\lambda,\gamma\right]$ can be obtained analytically by by solving the algebraic equation (\ref{eq16}) for $r_c$ 
using the Ferrari's method (for the details of the 
Ferrari's method and its use in classical algebra, see, e.g., \cite{kurosh72}).
Then,  from eq. (\ref{eq15}) one finds the flow velocity and
the sound speed for each solution  $r_c$. The critical 
space gradient of the flow velocity and
the sound speed evaluated on 
the acoustic horizon can then be obtained by applying l'Hospital's rule on the 
numerator and the denominator of $du/dr$ in eq. (\ref{eq14}), and then by substituting 
the value of $\left(du/dr\right)_{r_c}$ in the expression of 
$dc_s/dr$ in eq. (\ref{eq14}) on the acoustic horizon:
\begin{eqnarray}
\left(\frac{dc_{s}}{dr}\right)_{r_{c}}=
\frac{u_c}{r_{c}}\left(\frac{1-\gamma}{1+\gamma}\right)-
\sqrt{\frac{(1-\gamma)^2}{(1+\gamma)}
\left[\frac{u_c^2(1-3\gamma)}{4(1+\gamma)r_{c}^2}+
\frac{1}{4(r_{c}-1)^3}-\frac{3\lambda^2}{4r_{c}^4}\right]}  \\
\nonumber
\left(\frac{du}{dr}\right)_{r_{c}}=
\frac{u_c}{r_{c}}\left(\frac{1-\gamma}{1+\gamma}\right)-
\sqrt{\frac{u^2(1-3\gamma)}{r_{c}^2(1+\gamma)^2}+
\frac{1}{(1+\gamma)(r_{c}-1)^3}-\frac{3\lambda^2}{r_{c}^4(1+\gamma)}} 
\label{eq17}
\end{eqnarray}
Note that the quantities $r_c$,
$c_s$, $dc_s/dr$ and $du/dr$ evaluated at the acoustic horizon 
are expressed in terms of 
elementary functions  of
${\cal E}$, $\lambda$, and $\gamma$. Hence,  the 
surface gravity $\kappa$ can be calculated {\it analytically} 
as a function of $\left[{\cal E},\lambda,\gamma\right]$ since 
\begin{equation}
\kappa_{\text{\tiny CH}}=\zeta_{\text{\tiny CH}}\left(r,c_s,\frac{dc_s}{dr},\frac{du}{dr}\right)_{r_c}
\label{eq18}
\end{equation}
as is obvious from eq. (\ref{anal10}).
\subsubsection{Conical Model}
For conical flow, the space gradient of $c_s$ and $u$ can be obtained as
\begin{eqnarray}
\frac{dc_{s}}{dr}= \frac{c_{s}(1-\gamma)}{2} \left[ \frac{1}{u} \frac{du}{dr} + \frac{2}{r} \right] \\
\nonumber
\frac{du}{dr}= \frac{u \left[  \frac{2c_{s}^2}{r}+\frac{\lambda^2}{r^3} - 
\frac{1}{2(r-1)^2}  \right] }{(u^2-c_{s}^2)} 
\label{eq19}
\end{eqnarray}
Hence the critical point conditions becomes
\begin{equation}
\left(c_{s}\right)_{r_c}= \left(u\right)_{r_c} = 
\sqrt{\left[ \frac{r_{c}}{4(r_{c}-1)^{2}} - \frac{\lambda^{2}}{2r_{c}^2} \right]_{r_c}}
\label{eq20}
\end{equation}
The corresponding fourth degree polynomial in $r_c$ can be expressed as
\begin{eqnarray}
r_{c}^4+\Gamma_{1}r_{c}^3+\Gamma_{2}r_{c}^2+\Gamma_{3}r_{c}+\Gamma_{4}=0 \\
\nonumber
{\rm where} \\
\nonumber
\Gamma_{1}=\frac{(3\gamma-5)-16\mathcal{E}(\gamma-1)}{8\mathcal{E}(\gamma-1)}, 
~\Gamma_{2}=\frac{2(\gamma-1)(2\mathcal{E}-1)-\lambda^2(\gamma-3)}{4\mathcal{E}(\gamma-1)} \\
\nonumber
\Gamma_{3}=\frac{\lambda^2(\gamma-3)}{2\mathcal{E}(\gamma-1)}, 
~\Gamma_{4}=\frac{\lambda^2(3-\gamma)}{4\mathcal{E}(\gamma-1)}
\label{eq21}
\end{eqnarray}
The critical gradient of the sound speed and the flow velocity
can be obtained as
\begin{eqnarray}
\left(\frac{dc_{s}}{dr}\right)_{r_{c}}=
\frac{2u_c}{r_{c}}\left(\frac{1-\gamma}{1+\gamma}\right)-
\sqrt{\frac{u_c(1-\gamma)^4}{r_{c}^2(1+\gamma)^2}-
\frac{\lambda^2(2-r_{c})(1-\gamma)^2}{2r_{c}^4(1+\gamma)}+
\frac{(1-\gamma)^2[r_{c}(3-2\gamma)+(2\gamma-1)]}{8r_{c}(1+\gamma)(r_{c}-1)^3}} \\
\nonumber
\left(\frac{du}{dr}\right)_{r_{c}}=
\frac{2u_c}{r_{c}}\left(\frac{1-\gamma}{1+\gamma}\right)-
\sqrt{\frac{4u_c^2}{r_{c}^2}\left(\frac{1-\gamma}{1+\gamma}\right)^2-
\frac{2\lambda^2(2-r_{c})}{(1+\gamma)r_{c}^4}+ 
\frac{r_{c}(3-2\gamma)+(2\gamma-1)}{2r_{c}(1+\gamma)(r_{c}-1)^3}}
\label{eq22}
\end{eqnarray}
Using eq. (\ref{eq20}) - (\ref{eq22}) the acoustic surface gravity for the conical flow 
\begin{equation}
\kappa_{\text{\tiny CM}}=\zeta_{\text{\tiny CM}}\left(r,c_s,\frac{dc_s}{dr},\frac{du}{dr}\right)_{r_c}
\label{eq23}
\end{equation}
can thus be calculated analytically as a function of $\left[{\cal E},\lambda,\gamma\right]$.
 
\subsubsection{Flow in hudrostatic equilibrium in vertical direction}
For flow in hydrostatic equilibrium in the vertical direction, the velocity gradients and the 
corresponding critical point conditions become
\begin{eqnarray}
\frac{dc_{s}}{dr}=c_{s}\left(\frac{\gamma-1}{\gamma+1}\right) 
\left[\frac{-1}{u}\frac{du}{dr}-\frac{5r-3}{2r(r-1)}\right]\\
\nonumber
\frac{du}{dr}=u(\gamma+1)\frac{\left[\frac{\lambda^2}{r^3}-\frac{1}{2(r-1)^2}+
\frac{c_{s}^2(5r-3)}{r(r-1)(\gamma+1)}\right]}{u^2(\gamma+1)-2c_{s}^2}
\label{eq24}
\end{eqnarray}
\begin{equation}
\sqrt{\frac{2}{1+\gamma}}\left(c_s\right)_{r_c}=
\left(u\right)_{r_c}=
\sqrt{\frac{2\left[\Phi'(r_c)-\frac{\lambda^2}{r_c^3}\right]}{\left[\frac{3}{r_c}-\frac{\Phi''(r_c)}{\Phi'(r_c)}\right]}}
\label{eq25}
\end{equation}
The corresponding fourth degree polynomial in $r_c$ can be expressed as
\begin{eqnarray}
r_{c}^4+\Gamma_{1}r_{c}^3+\Gamma_{2}r_{c}^2+\Gamma_{3}r_c+\Gamma_4=0\\
\nonumber
{\rm where} \\
\nonumber
\Gamma_{1}=\frac{5-16\mathcal{E}-\frac{2\gamma}{\gamma-1}}{10\mathcal{E}},
~\Gamma_{2}=\frac{6\mathcal{E}-3+\frac{\gamma-5}{\gamma-1}\lambda^2}{10\mathcal{E}}\\
\nonumber
\Gamma_{3}=\frac{8\lambda^2}{10(\gamma-1)\mathcal{E}},
~\Gamma_4=\frac{(\gamma+3)\lambda^2}{10(\gamma-1)\mathcal{E}}
\label{eq26}
\end{eqnarray}
The critical gradient of the flow velocity can be found as
\begin{equation}
\left(\frac{du}{dr}\right)_{r_{c}}= \frac{-\beta-\sqrt{\beta^2-4\alpha\delta}}{2\alpha}
\label{eq27}
\end{equation}
where 
\begin{eqnarray}
\alpha=4\gamma \nonumber  \\
\beta=\frac{2(\gamma-1)(5r_{c}-3)u_c}{r_{c}(r_{c}-1)}  \nonumber  \\ 
\delta=\frac{\lambda^2(\gamma+1)(r_{c}-2)}{r_{c}^4(r_{c}-1)}-\frac{\gamma+1}{2r_{c}(r_{c}-1)^3}+\frac{(5r_{c}-3)(\gamma-1)}{2r_{c}(r_{c}-1)^3}  \nonumber  \\
-\frac{\lambda^2(5r_{c}-3)(\gamma-1)}{r_{c}^4(r_{c}-1)}-\frac{5(\gamma+1)}{2(r_{c}-1)^2(5r_{c}-3)}+\frac{5\lambda^2(\gamma+1)}{r_{c}^3(5r_{c}-3)}=0
\label{eq28}
\end{eqnarray}
Hence the critical gradient of the speed of sound can be found as
\begin{equation}
\left(\frac{dc_{s}}{dr}\right)_{r_c}=
\left({\sqrt{\frac{(\gamma+1)(r_c^3-2(r_c-1)^2\lambda^2)}{2r_c^2(r_c-1)(5r_c-3)}}}\right)\left(\frac{\gamma-1}{\gamma+1}\right) 
\left[\frac{-1}{u_c}\left(\frac{du}{dr}\right)_{r_c}-\frac{5r_c-3}{2r_c(r_c-1)}\right]
\label{eq29}
\end{equation}
where $\left(\frac{du}{dr}\right)_{r_c}$ is to be substituted from 
eq. (\ref{eq27}). 

\subsection{Isothermal accretion}
For isothermal flow under the influence of the Paczy\'nski and Wiita (1980) potential, 
specific energy does not remain one of the first integrals of motion 
any more. The mass accretion rate, however, still remains a constant of motion. Since the 
temperature is constant, the value of the isothermal sound speed 
$c_s=\sqrt{\kappa_B/(\mu m_H)}{T}^\frac{1}{2}$ 
is position independent 
and hence $dc_s/dr=0$ identically. Mach number profile for the isothermal 
accretion is thus found to be a scaled down version of the dynamical velocity profile. The 
stationary solution is completely characterized by two parameters 
$\left[T,\lambda\right]$, $T$ being the isothermal flow temperature. 
 
\subsubsection{Accretion flow with constant thickness}
For constant thickness flow, the velocity gradient 
\begin{equation}
\frac{du}{dr}=\frac{u\left[\frac{c_{s}^2}{r}+\frac{\lambda^2}{r^3}-
\frac{1}{2(r-1)^2}\right]}{(u^2-c_{s}^2)}
\label{eq30}
\end{equation}
provides the critical point condition as 
\begin{equation}
\left(u\right)_{r_c}=\left(c_s\right)_{r_c}=
\sqrt{\kappa_B/(\mu m_H)}{T}^\frac{1}{2}=
\sqrt{\left[\frac{r_{c}}{2(r_{c}-1)^2}-\frac{\lambda^2}{r_{c}^2}\right]}
\label{eq31}
\end{equation}
the corresponding fourth degree polynomial is
\begin{eqnarray}
r_{c}^4+\Gamma_{1}r_{c}^3+\Gamma_{2}r_{c}^2+\Gamma_{3}r_{c}+\Gamma_{4}=0\\
\nonumber
{\rm where}\\
\nonumber
\Gamma_{1}=-2-\frac{1}{2c_{s}^2},
\Gamma_{2}=1+\frac{\lambda^2}{c_{s}^2},
\Gamma_{3}=\frac{-2\lambda^2}{c_{s}^2},
\Gamma_{4}=\frac{\lambda^2}{c_{s}^2}
\label{eq32}
\end{eqnarray}
The critical flow velocity gradient becomes
\begin{equation}
\left(\frac{du}{dr}\right)_{r_{c}}=\sqrt{\frac{r_{c}+1}{4r_{c}(r_{c}-1)^3}-\frac{\lambda^2}{r_{c}^4}} 
\label{eq33}
\end{equation}
Although both $c_s$ and $du/dr$ evaluated at the acoustic horizon depend only 
on the angular momentum of the flow and not on the flow temperature, the location 
of the acoustic horizon itself (the critical point $r_c$) is a function of both 
$T$ and $\lambda$, hence 
\begin{equation}
\kappa^{\rm iso}_{\text{\tiny CH}}=\zeta_{\text{\tiny CH}}^{\rm iso}\left(r,c_s,\frac{du}{dr}\right)_{r_c}
\label{eq34}
\end{equation} 
can be calculated {\it analytically} (for all three different flow 
models considered here, as we will see in subsequent sections) as a function of
only two accretion parameters $\left[T,\lambda\right]$. 

\subsubsection{Conical Model}
For conical flow the velocity gradient
\begin{equation}
\frac{du}{dr}=\frac{u\left[\frac{2c_{s}^2}{r}+\frac{\lambda^2}{r^3}-
\frac{1}{2(r-1)^2}\right]}{(u^2-c_{s}^2)}
\label{eq35}
\end{equation}
provides the critical point condition as
\begin{equation}
\left(u\right)_{r_c}=\left(c_s\right)_{r_c}=
\sqrt{\kappa_B/(\mu m_H)}{T}^\frac{1}{2}=
\sqrt{\left[\frac{r_{c}}{4(r_{c}-1)^2}-\frac{\lambda^2}{r_{c}^2}\right]}
\label{eq36}
\end{equation}
The corresponding polynomial in $r_c$ becomes
\begin{eqnarray}
r_{c}^4+\Gamma_{1}r_{c}^3+\Gamma_{2}r_{c}^2+\Gamma_{3}r_{c}+\Gamma_{4}=0\\
\nonumber
{\rm where}\\
\nonumber
\Gamma_{1}=-2-\frac{1}{4c_{s}^2},
\Gamma_{2}=1+\frac{\lambda^2}{2c_{s}^2},
\Gamma_{3}=\frac{-\lambda^2}{c_{s}^2},
\Gamma_{4}=\frac{\lambda^2}{2c_{s}^2}
\label{eq37}
\end{eqnarray}
and the critical gradient of the dynamical velocity is thus
\begin{equation}
\left(\frac{du}{dr}\right)_{r_{c}}=\sqrt{\frac{r_{c}+1}{4r_{c}(r_{c}-1)^3}-\frac{\lambda^2}{r_{c}^4}}
\label{eq38}
\end{equation}
which is identical to that obtained for a  flow with constant thickness, see eq. (\ref{eq33}).
The corresponding acoustic surface gravity 
\begin{equation}
\kappa^{\rm iso}_{\text{\tiny CM}}=\zeta_{\text{\tiny CM}}^{\rm iso}\left(r,c_s,\frac{du}{dr}\right)_{r_c}
\label{eq39}
\end{equation}
can thus be calculated analytically as a function of $\left[T,\lambda\right]$. 
\subsubsection{Accretion in hydrostatic equilibrium in vertical direction}
For flow in hydrostatic equilibrium in the vertical direction, the corresponding 
quantities are
\begin{equation}
\frac{du}{dr}=u\frac{\left[\frac{c_{s}^2(5r-3)}{2r(r-1)}+
\frac{\lambda^2}{r^3}-\frac{1}{2(r-1)^2}\right]}{(u^2-c_{s}^2)}
\label{eq40}
\end{equation}
\begin{equation}
\left(u\right)_{r_c}=\left(c_s\right)_{r_c}=
\sqrt{\kappa_B/(\mu m_H)}{T}^\frac{1}{2}=
\sqrt{\frac{2r_{c}(r_{c}-1)}{(5r_{c}-3)}\left[\frac{1}{2(r_{c}-1)^2}-\frac{\lambda^2}{r_{c}^3}\right]}
\label{eq41}
\end{equation}
\begin{eqnarray}
r_{c}^4+\Gamma_{1}r_{c}^3+\Gamma_{2}r_{c}^2+\Gamma_{3}r_{c}+\Gamma_{4}=0\\
\nonumber
{\rm where} \\
\nonumber
\Gamma_{1}=-\frac{1+8c_{s}^2}{5c_{s}^2},
\Gamma_{2}=\frac{3c_{s}^2+2\lambda^2}{5c_{s}^2},
\Gamma_{3}=\frac{-4\lambda^2}{5c_{s}^2},
\Gamma_{4}=\frac{2\lambda^2}{5c_{s}^2}
\label{eq42}
\end{eqnarray}
\begin{equation}
\left(\frac{du}{dr}\right)_{r_{c}}=
\sqrt{\frac{1}{2r_{c}(r_{c}-1)(5r_{c}-3)}} \left[\frac{5r_{c}^2-3}{2(r_{c}-1)^2}-
\frac{2\lambda^2(5r_{c}^2-9r_{c}+3)}{r_{c}^3}\right]^{\frac{1}{2}}
\label{eq43}
\end{equation}
and the corresponding acoustic surface gravity
\begin{equation}
\kappa^{\rm iso}_{\text{\tiny VE}}=\zeta_{\text{\tiny VE}}^{\rm iso}\left(r,c_s,\frac{du}{dr}\right)_{r_c}
\label{eq44}
\end{equation}
can be evaluated accordingly. 

\section{Dependence of the acoustic surface gravity on the flow geometry
and initial boundary conditions}
\label{result}
\noindent
\begin{figure}[b]
\centering
\includegraphics[width=0.7\textwidth]{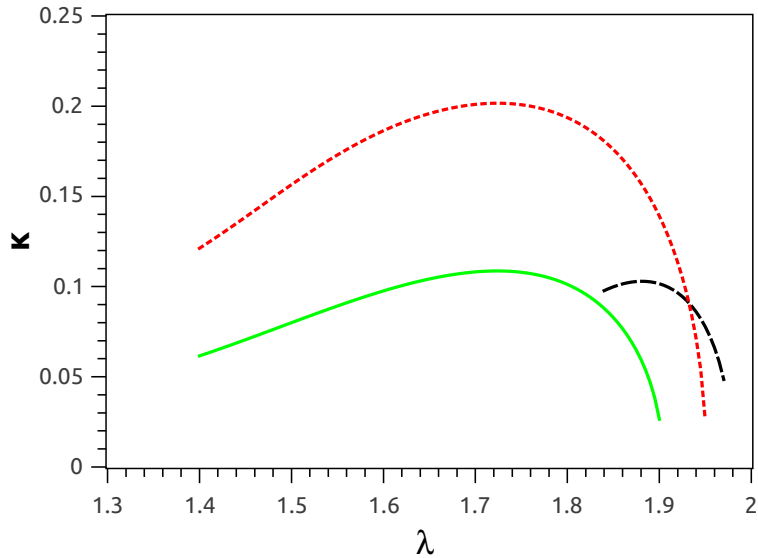}
\caption{Variation of the acoustic surface gravity with the specific angular momentum of the 
flow for monotransonic polytropic accretion characterized by $\mathcal{E}=0.06,\;\; \gamma=1.333$ 
for three different flow geometries - namely, for flow in hydrostatic equilibrium 
in the vertical direction (solid green line), constant thickness flow (long dashed black line),
and for the conical model (dotted red line).}
\label{lambadi}
\end{figure}

\begin{figure}[t]
\centering
\includegraphics[width=0.7\textwidth]{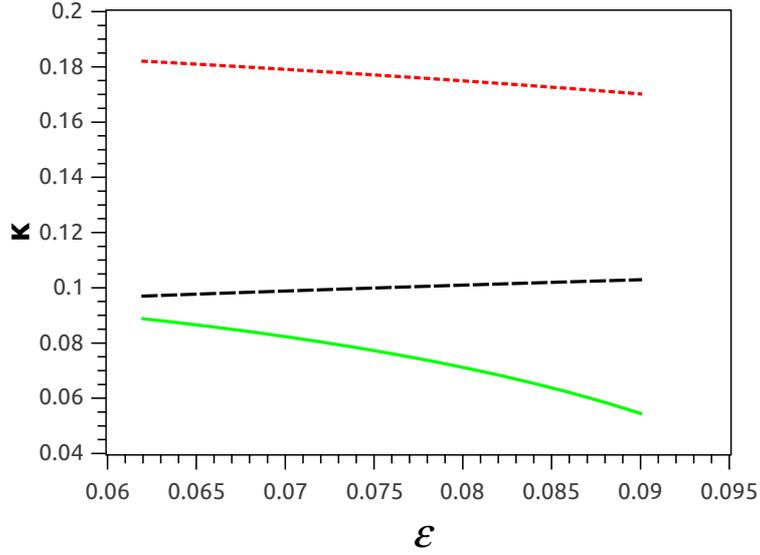}
\caption{Variation of the acoustic surface gravity with the specific energy of the 
flow for monotransonic polytropic accretion characterized by $\lambda=1.835,\;\; \gamma=1.333$
for three different flow geometries - namely, for flow in hydrostatic equilibrium 
in the vertical direction (solid green line), constant thickness flow (long dashed black line),
and for the conical model (dotted red line). Only the common range of specific energy for which all 
the flow configurations produce adiabatic monotransonic flow has been considered in the figure, see
text for further detail.}
\label{Enadi}
\end{figure}

\begin{figure}[t]
\centering
\includegraphics[width=0.7\textwidth]{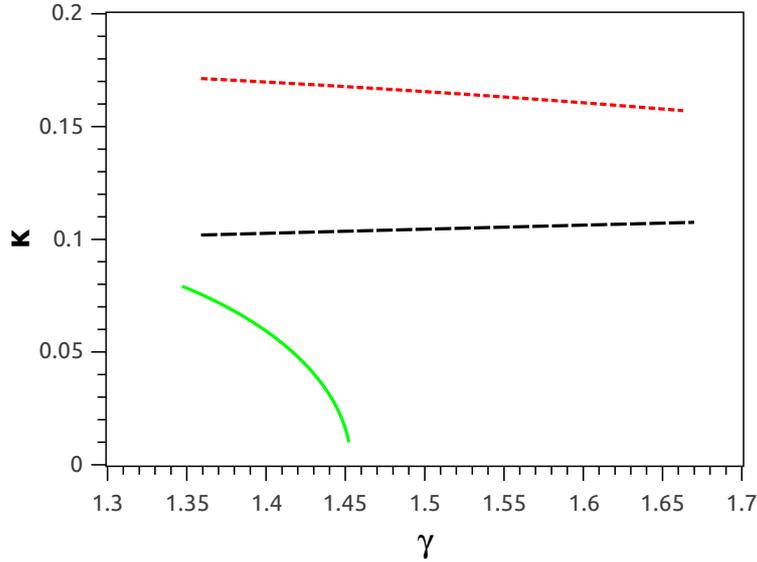}
\caption{Variation of the acoustic surface gravity with the specific energy of the          
flow for monotransonic polytropic accretion characterized by $\lambda=1.85,\;\; \mathcal{E}=0.06$
for three different flow geometries - namely, for flow in hydrostatic equilibrium
in the vertical direction (solid green line), constant thickness flow (long dashed black line),
and for the conical model (dotted red line).}
\label{gamadi}
\end{figure}

\begin{figure}[ht]
\centering 
\includegraphics[width=0.6\textwidth]{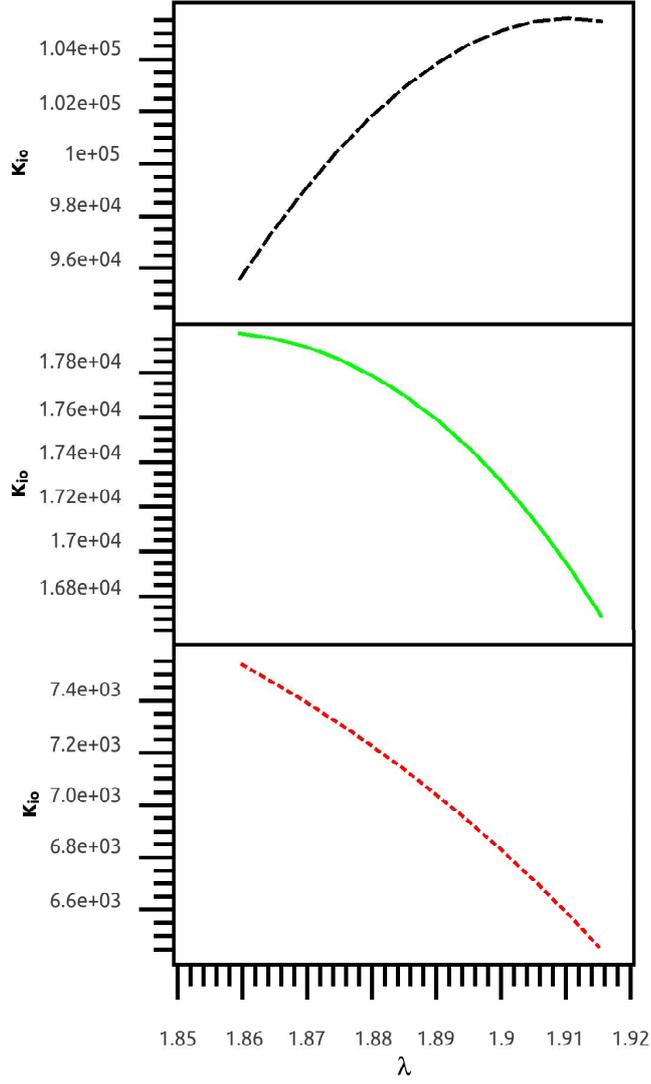}
\caption{Variation of the ratio of the acoustic surface gravity evaluated at the inner acoustic horizon to the 
acoustic surface gravity evaluated at the outer acoustic horizon with the range of the specific angular 
momentum (for a fixed value of $\left[{\cal E}=0.0002,\gamma=4/3\right]$) 
of the multi-transonic flow for which a stationary shock may form. }
\label{kio}
\end{figure} 

\begin{figure}[ht]
\centering 
\includegraphics[width=0.7\textwidth]{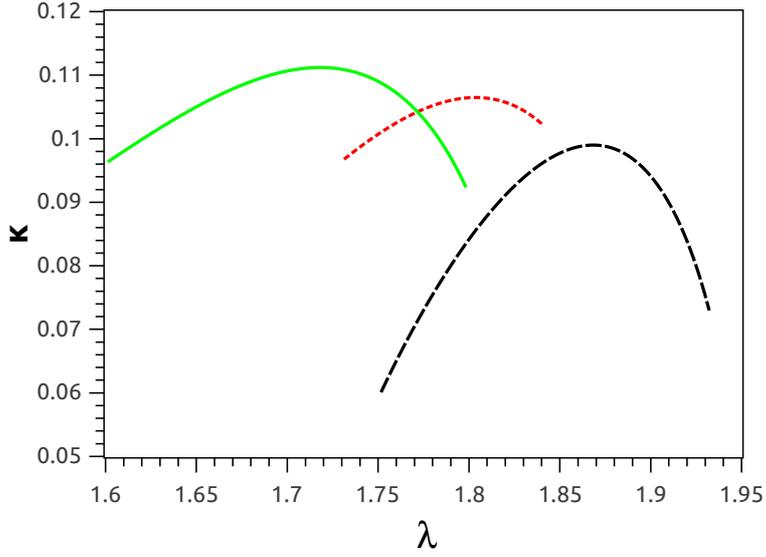}
\caption{Variation of the acoustic surface gravity with the specific angular momentum of the
flow for monotransonic isothermal accretion characterized by the isothermal flow temperature $T_{10}=22$
for three different flow geometries - namely, for flow in hydrostatic equilibrium
in the vertical direction (solid green line), constant thickness flow (long dashed black line),
and for the conical model (dotted red line).}
\label{lamiso}
\end{figure}

\begin{figure}[hb]
\centering
\includegraphics[width=0.7\textwidth]{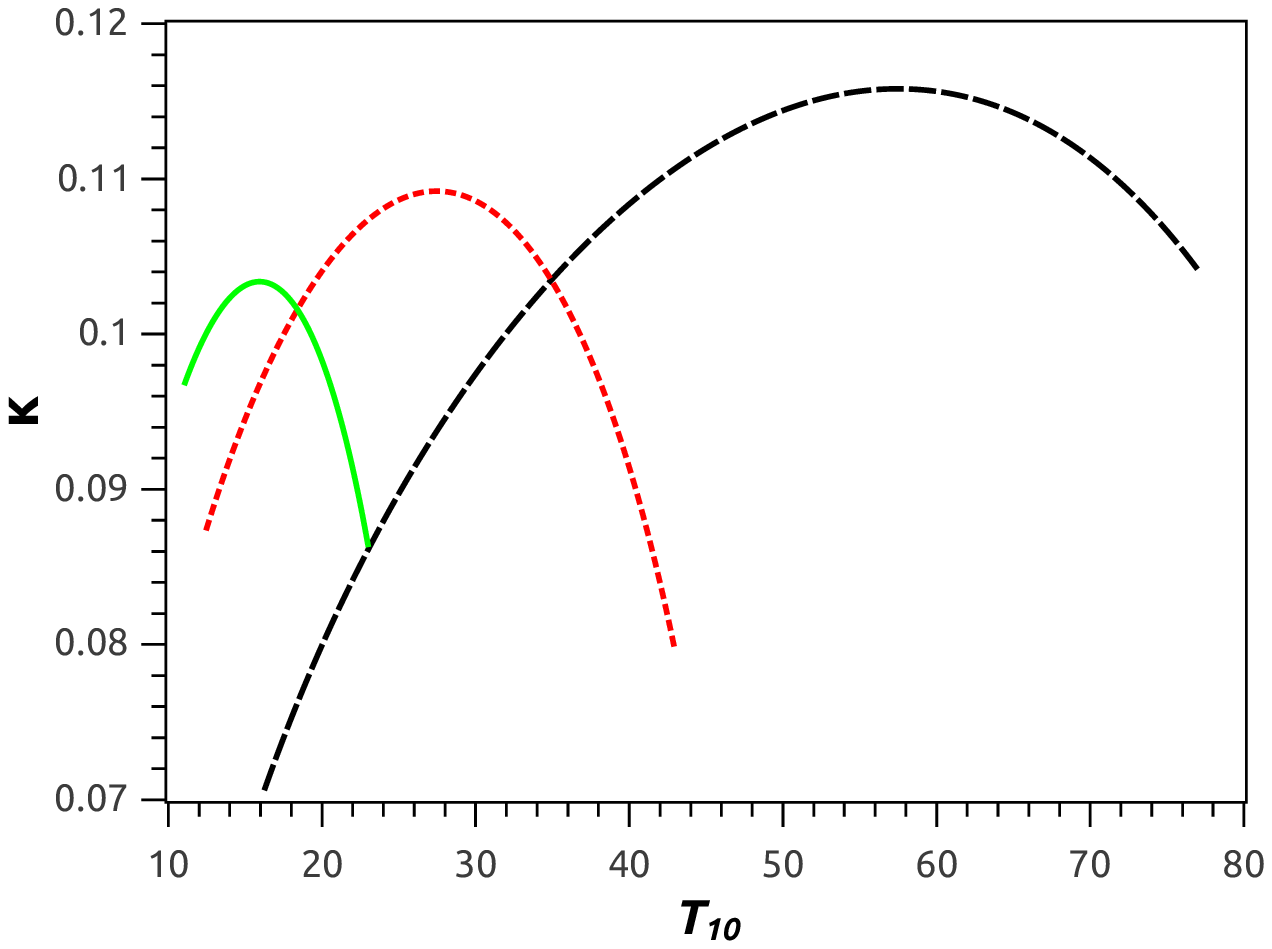}
\caption{Variation of the acoustic surface gravity with the isothermal flow temperature of the
flow for monotransonic isothermal accretion characterized by$\lambda=1.8$
for three different flow geometries - namely, for flow in hydrostatic equilibrium
in the vertical direction (solid green line), constant thickness flow (long dashed black line),
and for the conical model (dotted red line).}
\label{Tiso}
\end{figure}

Figure~\ref{lambadi} shows the variation of the acoustic surface gravity $\kappa$ as a
function of the specific angular momentum $\lambda$ for monotransonic polytropic accretion.
The specific energy ${\cal E}$ and the polytropic index $\gamma$ have been 
kept constant at values $\left[{\cal E}=0.06,\gamma=1.333\right]$. The range 
of $\lambda$ for which $\kappa$ has been calculated for a particular 
flow model constructs a subset in the $\left[{\cal E},\lambda,\gamma\right]$
parameter space for which the representative flow models produces 
mono-transonic accretion for a fixed set of values of 
$\left[{\cal E}\right]$. Alternative ranges for $\lambda$ for 
other similar subsets of $\left[{\cal E},\lambda,\gamma\right]$ 
parameter space may also be considered to study the `$\kappa - \lambda$' 
profile for other fixed values of $\left[{\cal E},\gamma\right]$. The solid
line (green coloured in the online version) represents the 
$`\kappa$ - $\lambda$' variation profile for 
mono-transonic accretion in vertical equilibrium (VE), whereas the dotted 
(red coloured in the online version) and the long dashed (black 
coloured in the online version) curves represents such dependence 
for flow with constant height (CH) as well as for the conical 
flow (CM) respectively. 

It is usually observed that $\kappa$ varies with $\lambda$ non linearly and 
non monotonically. For relatively lower values of the specific 
angular momentum, the acoustic surface gravity correlates with the 
specific angular momentum and attains a peak characterized by 
an unique value of the specific angular momentum denoted by 
$\lambda_{\rm max}$, and subsequently falls of with $\lambda$ 
for $\lambda>\lambda_{\rm max}$. $\lambda_{\rm max}$ is different 
for different flow models and one observes that 
\begin{equation} 
\lambda^{\rm CM}_{\rm max} > \lambda^{\rm VE}_{\rm max} >
\lambda^{\rm CH}_{\rm max}
\label{result1}
\end{equation}
For a set of fixed values of $\left[{\cal E},\lambda\right]$,
$\lambda_{\rm max}$ for a particular flow model can be calculated 
completely analytically. We illustrate such procedure for 
constant thickness flow. Similar procedure may be followed to 
calculate $\lambda_{\rm max}$ for two other flow geometries. 

For constant thickness flow, eq. (\ref{eq18}) provides the dependence 
of $\kappa$ on the critical points, the acoustic speed and its space
gradient, and on the space gradient of the 
flow velocity itself, everything evaluated at the acoustic horizon 
(the critical point). $\left(c_s\right)_{r_c}$, 
$\left(dc_s/dr\right)_{r_c}$ and $\left(du/dr\right)_{r_c}$ can be 
expressed as a function of $r_c$ and $\left[{\cal E},\lambda,\gamma\right]$ 
using eq. (\ref{eq15}) and eq. (\ref{eq17}), respectively. The critical point 
$r_c$ itself can be computed in terms of 
$\left[{\cal E},\lambda,\gamma\right]$ by solving the polynomial in $r_c$ as represented through eq. (\ref{eq16}). 
Hence $\kappa$ for the constant 
thickness flow, can be specified in terms of
$\left[{\cal E},\lambda,\gamma\right]$. Analytical expression for 
$\kappa_{\rm CH}{\equiv}\kappa_{\rm CH}\left[{\cal E},\lambda,\gamma\right]$

can thus be maximized with respect to the specific angular momentum and the 
corresponding $\lambda_{\rm max}$ can thus be obtained. 

From figure~\ref{lambadi}, one should note that for the common range of the 
specific angular momentum for which all three flow models will 
produce mono transonic flow for a fixed set of value of 
$\left[{\cal E},\lambda\right]$, does not allow to explore the 
complete non monotonic `$\kappa$ - $\lambda$' profile for all the three
flow models simultaneously. For example, for the common range of $\lambda$
as shown in the figure for which all three flow geometries produces the 
monotransonic accretion, $\kappa$ for constant height flow as well as for 
conical model accretion will anti correlates with $\lambda$, whereas
for full range of allowed $\lambda$ to form mono transonic accretion at individual level,
`$\kappa$ - $\lambda$' profile for both the aforementioned flows 
exhibits a maximum. 


Similar features are observed for the `$\kappa$ - ${\cal E}$' profile, which has been 
shown in the figure~\ref{Enadi}, where the $\kappa$  vs ${\cal E}$ variation (for a fixed set of 
$\left[\lambda,\gamma\right]$) for constant height flow apparently shows that 
$\kappa_{\rm CH}$ correlates with ${\cal E}$, whereas $\kappa_{\rm CM}$ 
and $\kappa_{\rm VE}$ anti-correlates with ${\cal E}$. Such trend, however, does not provide the 
complete information about the `$\kappa$ - ${\cal E}$' profile in general since the range of
${\cal E}$ for which the figure is drawn is actually taken from the common 
region of $\left[{\cal E},\lambda,\gamma\right]$ space for which all three flow geometries
produces mono transonic accretion for a fixed value of $\left[\lambda,\gamma\right]$. If one 
allows $\kappa$ to vary with ${\cal E}$ for the entire range of the specific energy for which 
a particular flow model provides the mono transonic accretion for a fixed value of 
$\left[\lambda,\gamma\right]$, $\kappa$ vs ${\cal E}$ profile would have a non monotonic behaviour 
with a corresponding ${\cal E}_{\rm max}$ separately for {\it every} flow configuration. 
The corresponding  ${\cal E}_{\rm max}$ for every flow model could then be estimated 
by maximizing the expression for the acoustic surface gravity with respect to 
${\cal E}$ by keeping $\left[\lambda,\gamma\right]$ constant. One thus understands that 
the `anti correlating' $\kappa$ vs ${\cal E}$ profile for the conical flow 
(represented by the dotted red curve) as well as for flow in the hydrostatic equilibrium 
along the vertical direction (solid green curve), respectively, are thus the post-peak 
(${\cal E} > {\cal E}_{\rm max}$) descending part of the complete non monotonic 
`$\kappa$ - ${\cal E}$' profiles for the corresponding flow configuration. Similarly, 
the `correlating' $\kappa$ - ${\cal E}$ profile for the constant height flow (represented 
by the long dashed black curve) is actually the pre-peak (${\cal E} < {\cal E}_{\rm max}$) 
ascending part of the complete non monotonic $\kappa$ - ${\cal E}$ profile for the 
corresponding flow geometry. Hence ${\cal E}^{\rm CH}_{\rm max}$ is largest among all
the values of ${\cal E}_{\rm max}$ corresponding to all three different flow configurations. 

Figure~\ref{gamadi} shows the $\kappa$ - $\gamma$ profile (for a fixed set of $\left[{\cal E},\lambda\right]$).
It is obvious from the figure that  ${\gamma}^{\rm CH}_{\rm max}$ has the largest value
among all three values of ${\gamma}_{\rm max}$ corresponding to three different flow 
configurations. ${\cal E}_{\rm max}$ and $\gamma_{\rm max}$ for any particular model can also 
be estimated by maximizing $\kappa$ with respect to the respective parameters. 

One thus understands that for monotransonic adiabatic accretion characterized by a 
fixed set of values of $\left[{\cal E},\lambda,\gamma\right]$, the analogue surface 
gravity for three different flow configurations exhibit the following trend
\begin{equation} 
\kappa_{\text{\tiny CM}}^{\rm adia}~>~\kappa_{\text{\tiny VE}}^{\rm adia}~>~\kappa_{\text{\tiny CH}}^{\rm adia}
\label{result3}
\end{equation}
Hence for the adiabatic mono transonic accretion, 
flow in conical shape produces the largest value of the analogue Hawking like temperature
whereas the flow with constant thickness provides the lowest value of $T_{AH}$ for the 
same set of initial boundary conditions. 

As has been mentioned in section \ref{polytropic_accretion}, multi-transonic accretion with 
stationary shock may be realized for all three different flow configurations 
for adiabatic as well as for isothermal accretion,
studied here in this work. For such flow topologies, two black hole type 
acoustic horizons form at the inner and the outer saddle type sonic points. The 
corresponding acoustic surface gravity $\kappa_{in}$ and $\kappa_{out}$ can be 
evaluated at the inner and the outer sonic points respectively. We define 
\begin{equation}
\kappa_{io}=\frac{\kappa_{in}}{\kappa_{out}}
\label{result4}
\end{equation}
It has been observed that the overall $\left(\kappa_{in} - \left[{\cal E},\lambda,\gamma\right]\right)$
as well as the $\left(\kappa_{out} - \left[{\cal E},\lambda,\gamma\right]\right)$ profiles are 
qualitatively similar with the $\left(\kappa - \left[{\cal E},\lambda,\gamma\right]\right)$ 
profile for all three flow configurations, where $\kappa$ is the acoustic surface gravity 
evaluated for the mono transonic accretion. For all three flow geometries considered in this 
work, the value of $\kappa_{out}$ is, however, several orders of magnitude
less than that of the corresponding $\kappa_{in}$ for same set of initial boundary conditions. 
This is because the outer acoustic horizons form at a large distance away 
from the black hole event horizon compared to the location of the inner acoustic horizon 
with respect to the black hole event horizon. For a typical set of the initial boundary 
conditions, the inner acoustic horizon may form 1.5 - 5 Schwarzschild radius away from the 
black hole event horizon whereas the outer acoustic horizon may be located at $10^3$ - $10^6$ 
Schwarzschild radius away, or even more, from the black hole event horizon. Weak 
gravity at such a large distance (where the outer sonic horizons form) restrict the 
acoustic surface gravity to possess such a small numerical value. Same argument applies 
for the analogue Hawking temperatures evaluated at the inner and the outer acoustic horizons 
as well. 

In figure~\ref{kio}, we plot $\kappa_{io}$ as a function of $\lambda$ for the constant 
thickness flow (uppermost panel), flow in hydrostatic equilibrium along the vertical 
direction (mid panel) and for the conical flow (lowermost panel). The range of $\lambda$ 
used in this figure corresponds to the common value of the specific angular 
momentum for which shock forms for multi-transonic accretion in all three 
flow geometries for a fixed value of $\left[{\cal E},\lambda\right]$. The common range of 
$\lambda$ is chosen in such a way so that the inner acoustic horizon forms at a distance 
larger than two Schwarzschild radius from the black hole event horizon. Since we 
use pseudo Schwarzschild potentials which are relatively less reliable in simulating the 
general relativistic space time extremely close to the black hole event horizon, we 
prefer to confine our attention to the mono transonic as well as the multi transonic flow 
topologies for which the acoustic horizon does not form at a very close proximity to the 
black hole event horizon. From the figure it is evident that 
\begin{equation} 
\lambda^{\rm CM}_{\rm max} > \lambda^{\rm VE}_{\rm max} >
\lambda^{\rm CH}_{\rm max}
\label{result5}
\end{equation}
where $\lambda_{\rm max}$ in eq. (\ref{result5}) is the value of $\lambda$ for which 
the non monotonic $\kappa_{io}$ - $\lambda$ profile attains its maximum. 
Interestingly enough, eq. (\ref{result1}) is identical with eq. (\ref{result5}), indicating the 
fact that the dependence of the acoustic surface gravity on initial boundary 
conditions is similar for both the mono as well as for the multi-transonic shocked 
accretion in all three flow geometries considered here in this work.  
Once again, the $\kappa_{io}$ - $\lambda$ dependence does not exhibit the 
peaked non monotonic profile since the common range corresponding to the 
shock forming $\lambda$ is not sufficient to provide the required span for the 
$\kappa_{io}$ - $\lambda$ variation for any particular flow geometry for the 
entire range of $\lambda$ for which the corresponding flow model produces shocked 
multi-transonic accretion for a fixed value of $\left[{\cal E},\lambda\right]$. 

Figure~\ref{lamiso} demonstrates the dependence of the acoustic surface gravity on specific angular momentum
for mono transonic isothermal accretion for three different flow models. It is observed that 
\begin{equation} 
\lambda^{\rm CM}_{\rm max} > \lambda^{\rm VE}_{\rm max} >
\lambda^{\rm CH}_{\rm max}
\label{result6}
\end{equation}
In figure~\ref{Tiso}, we plot the dependence of $\kappa$ on the isothermal flow temperature $T$ 
(scaled by a factor of $10^{10}$ degree Kelvin - $T_{10}{\equiv}T{\times}{10^{-10}}$). 
We obtain
\begin{equation} 
T^{\rm CM}_{\rm max} > T^{\rm VE}_{\rm max} >
T^{\rm CH}_{\rm max}
\label{result7}
\end{equation}
The value of $\lambda_{\rm max}$ and $T_{\rm max}$ for the monotransonic isothermal flow may be 
estimated for all three flow geometries in a way similar to what has been accomplished 
for the polytropic flow. 

\section{Discussion}
Axisymmetric accretion onto a non-rotating astrophysical black hole 
under the influence of pseudo-Schwarzschild potentials are natural example
of classical analogue systems found in the universe. The corresponding acoustic geometry 
may be studied for three different flow configurations, viz, accretion 
with constant flow thickness, the conical flow, and accretion disc in hydrostatic equilibrium 
in the vertical direction. For each background flow geometry, all together 
eight different configurations for the acoustic geometry - adiabatic as well as
isothermal accretion for four different pseudo-potentials, may be studied. For 
any specific pseudo potential, six different configurations of the 
acoustic flow geometry - adiabatic 
as well as isothermal accretion in three different flow geometries, may 
be studied. For any geometric configuration of the  adiabatic flow
under the influence of a particular potential,
 the three  initial parameters, viz., the specific energy ${\cal E}$, the 
specific angular momentum $\lambda$, and the adiabatic index of the flow
$\gamma$ can completely specify the corresponding acoustic geometry. Similarly 
for the isothermal flow in any potential and with any flow geometry, the two  initial parameters,
viz., the constant flow temperature $T$ and the specific angular momentum $\lambda$,
can completely specify the corresponding acoustic geometry.

Among six possible flow configurations for any particular pseudo potential
used, only the adiabatic flow in vertical equilibrium exhibits an 
`effective' sound speed which is a scaled version of the adiabatic sound 
speed with a $\gamma$ dependent scaling constant. The scaling constant becomes 
unity for isothermal flow. The reason is that for the 
accretion in vertical equilibrium the flow thickness is a function of the adiabatic sound speed, 
as well as a function of the space derivative of the pseudo-potential 
used, since the expression for such flow thickness is obtained 
by balancing the pressure gradient with the relevant component of the 
gravitational force. One should, however, bear in mind that the corresponding 
expressions for the flow thickness in all three flow geometries used are derived 
using a set of idealized assumptions. In principle, a more realistic derivation of 
the flow thickness may be worked out by employing the non-LTE 
radiative transfer (see~\cite{hh98, dh06} ) or
by taking recourse to the Grad-Shafranov equations for MHD flow (see~\cite{beskin97, 
bt05, beskin09} ).

For multi-transonic flow, two acoustic black holes are formed at 
two regular saddle type sonic points, whereas the acoustic white 
hope forms at the shock location, in agreement 
with the results obtained by 
\cite{blsv04}  and 
\cite{abd06} . The acoustic surface gravity 
is formally infinite for the acoustic white hole since the 
flow velocity as well as the sound speed changes discontinuously at 
the shock location, in agreement with the findings of 
\cite{lsv00} .

The surface gravity $\kappa$ (or $T_{\rm AH}$) profile obtained for
the multi-transonic
flow at the inner acoustic horizon is similar to the
$\kappa$ profile for mono-transonic flow. This indicates that
irrespective of the phase topology, the surface gravity
is basically determined by the physical proximity of the acoustic horizon to
the black hole event horizon. This is further supported by
the fact that irrespective of the flow geometry,
pseudo potentials or the equation of state
used to describe the accretion flow, the value of $\kappa$ at the outer acoustic
horizon (for multi-transonic flow) is much less than that
evaluated at the inner acoustic horizon.

For a fixed set of $\left[{\cal E},\lambda,\gamma\right]$
describing the adiabatic accretion as well as for a fixed set of 
$\left[T,\lambda\right]$ describing the isothermal accretion, 
the conical flow produces the largest whereas  the flow with constant thickness the smallest surface gravity and the analogue Hawking temperature. The accretion flow in the 
hydrostatic equilibrium in the vertical direction provides the 
value of $\kappa$ and $T_{\rm AH}$ in between the respective
values of $\kappa$ and $T_{\rm AH}$ for the 
conical and the constant thickness flow, respectively. 
The influence of the flow geometry in determining the acoustic surface gravity has thus been 
successfully investigated in this work. 

One of our major achievements is the computation of the acoustic surface 
gravity and the investigation of its dependence on various flow geometries and 
on various accretion parameters {\it completely 
analytically.}. However,
one should note that this has been possible 
owing to a specific feature of the chosen pseudo potential. With the 
Paczy\'nski \& Wiita (1980) potential $\Phi_1=-\frac{1}{2\left(r-1\right)}$,
the energy first integral for the adiabatic accretion (see eq. (\ref{eq13})
as well as the critical point condition for the isothermal flow 
(see eq. (\ref{eq17} - \ref{eq19})) can be recast into a fourth degree
polynomial in ${r_c}$ (equivalently, in ${r_h}$) in an exactly solvable form. This has not been possible with other 
pseudo potentials, for which such an 
exactly solvable polynomial in ${r_h}$ can not be constructed. 
With other potentials, however, the surface gravity $\kappa$ can still be studied  
as a function of various flow parameters and  
flow geometries, with the help of numerical 
 methods. Remarkably,
it has been demonstrated in the literature 
(\cite{abn96, das02} 
and references therein) that out of the four pseudo
Schwarzsculd potentials as shown in 
eq. (\ref{eq21}), the potential $\Phi_1$  mimics the 
Schwarzschild space time most efficiently in constructing the integral flow solution for transonic accretion.

However, even if one can not employ a complete analytical 
calculation and even if $\Phi_1$ is the most suitable 
potential to mimic a Schwarzschild space time, our generalized 
formalism for the evaluation of
$\left[u,c_s,dc_s/dr,du/dr\right]_{r_h}$ and the 
corresponding values of $\kappa$ and $T_{\rm AH}$ in terms 
of a general $\Phi$ is still important in the following sense.
There exists a possibility that a new form of $\Phi$ more effective than $\Phi_1$ will be suggested in the future in order  to
approximate the Schwarzschild space time in constructing the 
integral accretion solutions for the transonic flow. In such a case,
if a general model is 
capable of computing the acoustic surface gravity 
(and hence the analogue Hawking temperature) in a way presented in this work,
then this model 
will be able to readily accommodate that novel form 
of the pseudo-Schwarzschild potential with no need to significantly
 change the fundamental structure 
of the formulation and the solution scheme. In this case one need 
not worry about providing any new unique scheme 
valid exclusively only for a particular form of pseudo
potential.

The methodology developed in this paper
can also be used to construct the relevant acoustic geometry for the
equatorial slice of the accretion flow under the influence of 
various pseudo-Kerr potentials as well to study the variation of 
the surface gravity with the black hole spin. Such work is in 
progress and will be reported elsewhere. 

However, one  should be cautious when using the pseudo 
potentials because none of the potentials discussed 
here can be directly derived from the 
Einstein equations. These potentials are used to obtain more accurate correction 
terms over and above the pure Newtonian results. Hence any `radically new' 
results obtained using these potentials should be cross checked very carefully 
against general relativity. Besides, one should bear in mind that 
these potentials are not too reliable for modeling  the 
space  time in the  very close neighborhood of the event horizon  since 
strong gravity effects dominate in this region. Hence
our formalism may not be quite realistic if the acoustic horizon forms 
very close to the black hole event horizon. We thus 
consider only those initial boundary conditions for which ${r_h > 2}$ ensuring our findings to be trustworthy.

\acknowledgments{AC would like to acknowledge the kind hospitality
provided by HRI, Allahabad, India, under a visiting student
research programme. The visits of SN at HRI was partially supported
by astrophysics project under the XIth plan at HRI.
The work of TKD has been partially supported by a research grant
provided by S. N. Bose National Centre for Basic
Sciences, Kolkata, India, under a guest scientist (long term
sabbatical visiting professor) research
programme. The research of
TKD is partially funded by the astrophysics project under the XI th
plan at HRI. The work of NB has been supported by the Ministry of Science, Education and Sport of the Republic of Croatia under Contract No.~098-0982930-2864.}

\providecommand{\href}[2]{#2}\begingroup\raggedright\endgroup

\end{document}